\documentclass[epj]{svjour}
\usepackage{graphicx,color}
\usepackage{amsmath}
\usepackage{amssymb}
\usepackage{amsfonts} 
\begin{document}
\title{Measuring economic complexity of countries and products: which metric to use?}
\author{Manuel Sebastian Mariani\thanks{\emph{e-mail:} manuel.mariani@unifr.ch}, Alexandre Vidmer, Mat{\'u}{\v{s}} Medo, \and Yi-Cheng Zhang
}                     
%
%
\institute{Department of Physics, University of Fribourg, Chemin du Mus\'{e}e 3, CH-1700 Fribourg, Switzerland} 
\date{}
%
\abstract{
Evaluating the economies of countries and their relations with products in the global market is a central problem in
economics, with far-reaching implications to our theoretical understanding of the international
trade as well as to practical applications, such as policy making and financial investment planning.
The recent Economic Complexity approach
aims to quantify the competitiveness of countries and the quality of the exported products based on 
the empirical observation that the most competitive countries have diversified exports, whereas developing countries only
export few low quality products---typically those exported by many other countries.
Two different metrics, Fitness-Complexity and the Method of Reflections, have been proposed to measure country
and product score
in the Economic Complexity framework.
We use international trade data and a recent ranking evaluation measure 
to quantitatively compare the ability of the two metrics to rank countries and products according to their 
importance in the network. The results show that the Fitness-Complexity metric 
outperforms the Method of Reflections in both the ranking of products and the ranking of countries. 
We also investigate a generalization of the Fitness-Complexity metric and show that
it can produce improved rankings provided that the input data are reliable.
%
} 

\authorrunning{Mariani M. S., Vidmer A., Medo M., Zhang Y.-C.}
\titlerunning{Measuring economic complexity: which metric to use?}
\maketitle
\section{Introduction}
\label{intro}

Where does the wealth of nations come from? Which countries will economically grow in the future?
Unveiling the intangible factors driving economic success and growth is a long-standing problem which still has 
many open questions.
Classical economic theories, from Ricardo's theory \cite{ricardo1891principles} to more recent studies \cite{romer1990endogenous,grossman1991quality}, 
emphasize the importance of specialization of countries' production
on few high-quality products.
The Economic Complexity approach exploits the interconnectedness of the current global market and 
represents the international trade data as a bipartite network where countries are connected to the products that they export.
The starting point of the analysis is the empirical observation that the most competitive countries tend to diversify
their export basket, whereas developing countries are able to export only a few products, typically those exported by many other 
countries \cite{hidalgo2009building,tacchella2012new}. This finding contrasts the standard view that
the richest countries should specialize their production in economic niches
\cite{ricardo1891principles}.

How to best infer countries' and products' economic complexity from the structure of the country-product network is still debated.
There are two competing viewpoints: (1)  The Method of Reflections (MR) by Hidalgo and Hausmann \cite{hidalgo2009building}
defines country and product complexity through a set of linear iterative equations,
similarly to Google's PageRank \cite{brin1998anatomy};
(2) The Fitness-Complexity Method (FCM) by Tacchella et al. \cite{tacchella2012new} defines country fitness and product complexity
through a set of non-linear iterative equations.
The input of both methods is the adjacency matrix of the country-product network.
The MR has been applied to world trade data both by the original authors \cite{hausmann2014atlas}
and by others \cite{felipe2012product,poncet2013export,cheng2013hidden}.
It has been shown that the complexity index based on the MR scores -- called 
Economic Complexity Index (ECI) -- 
contributes
to the variance of countries' economic growth
significantly more than the existing governance, institutional quality, education quality and economic competitiveness indexes \cite{hausmann2014atlas}. 
By contrast, Refs. \cite{tacchella2012new,cristelli2013measuring,caldarelli2012network,tacchella2013economic,battiston2014metrics}
emphasize some negative aspects of the application of the MR to the world trade data.
This criticism is mostly motivated by the study of a few particular countries
and by the convergence of the iterations that define the MR scores toward an uniform fixed point \cite{cristelli2013measuring,caldarelli2012network}.
While the convergence issue can be fixed by a mathematical transformation of 
the MR variables \cite{hausmann2014atlas}, the case studies pointed out in Ref. \cite{cristelli2013measuring} indicate that the MR
underestimates the importance of highly diversified countries, such as China and India. 
Unlike the MR, 
the non-linear equations that define the FCM favor countries with a diversified export basket and penalize products that are exported
by poorly diversified countries \cite{tacchella2012new}.

However, a quantitative comparison of the quality of the rankings of countries and products by the two methods is still lacking. 
In this work, we fill this gap by applying both methods on the NBER-UN world trade dataset and by comparing the resulting rankings of countries and products.
To evaluate the rankings, we use a metric introduced in the context of ecological networks, the extinction area \cite{allesina2009googling}.
This metric reflects the idea that developed countries are fundamental for the presence of complex products in the world trade,
while non-complex products are crucial for the production of developing countries.
We find that the FCM clearly outperforms the MR in this respect.
This is in agreement with the findings on ecological networks \cite{dominguez2015ranking} where
the FCM has been shown to be the best candidate to rank
active and passive species, such as pollinators and plants, according to their importance and vulnerability, respectively.

While the non-linear equations of the FCM introduced in \cite{tacchella2012new} represent one the simplest mathematical
equations that favor countries with diversified exports and penalize products with a large number of exporter countries,
their possible generalizations have not been yet applied
to world trade data.
Here, we study a generalization of the original FCM where the dependence of product complexity on fitness of the countries that export
is governed by a tunable parameter $\gamma$.
We find that when the ability of the generalized FCM to rank nodes by importance can be improved by changing $\gamma$, 
the country and product rankings become more sensitive to noise
and changes more rapidly with time.
This discourages the use of the generalized metric when a significant level of noise is present in the data, as is the case for the world 
trade data \cite{battiston2014metrics}.

\section{Materials and methods}
\label{sec:1}

\subsection{Dataset and definition of the country-product network}

We use the NBER-UN dataset which has been cleaned and further described in \cite{feenstra2005world}.
We take into account the same list of $N=132$ countries described in \cite{hidalgo2007product}.
For products, we used the same cleaning procedure of Ref. \cite{vidmer2015prediction}: we removed
aggregate product categories and products with zero total export volume for a given year and
nonzero total export volume for the previous and the following years.
Products and countries with no entries after year 1993 have been removed as well.
After the cleaning procedure, the dataset consists of $M=723$ products. 
To decide if we consider country $i$ to be an exporter of product $\alpha$ or not, we use the Revealed Comparative
Advantage (RCA) \cite{balassa1965trade} which is defined as
\begin{equation}
RCA_{i\alpha}=\frac{e_{i\alpha}}{\sum_\beta e_{j\beta}} \Bigg/ \frac{\sum_j
e_{j\alpha}}{\sum_{j\beta}e_{j\beta}},
\end{equation}
where $e_{i\alpha}$ is the volume of product $\alpha$ that country $i$ exports measured in thousands of US dollars.
RCA characterizes the relative importance of a given export volume of a product
by a country in comparison with this product's exports by all other countries.
We use the bipartite network representation introduced in \cite{hidalgo2009building}, where two kinds of nodes represent countries and products, respectively.
All country-product pairs with RCA values above a threshold value--set to $1$ here--are consequently joined by links between the corresponding nodes in
the bipartite network.
We refer to the degree $d_{i}$ of country $i$ and to the degree $u_{\alpha}$ of product $\alpha$ as country $i$'s diversification and product $\alpha$'s 
ubiquity, respectively \cite{hidalgo2009building}.
The adjacency matrix $M_{i\alpha}$ of the resulting network is the only input of the metrics 
for economic complexity \cite{hidalgo2009building,tacchella2012new}.
Weighted metrics based on the fraction of the total export instead of $M_{i\alpha}$
provide useful information as well \cite{tacchella2012new} but are not studied in this work.
In the NBER dataset, the complete information about mutual exchanges is only available for a core group of $72$ countries.
For countries outside this core group, we know only their exchanges with countries inside the core group, but we have actually no information about
exchanges between them \cite{feenstra2005world}.
This incompleteness of data makes the robustness against noisy data a crucial element to evaluate the metrics (see Section \ref{robust}).
We studied also a restricted dataset that only contains the $72$ countries belonging to the core group.
We found that neglecting the countries outside the core group can lead to inconsistencies in the rankings due to the loss of information on
the exports from developing countries, as we will discuss in Section \ref{cleaning}.

\subsection{Method of reflections (MR)}
\label{sec:2}

The Economic Complexity Index by Hidalgo and Hausmann is built on the method of reflections \cite{hidalgo2009building,hausmann2014atlas}.
This method defines the $n$-th order country scores $\{d_{i}^{(n)}\}$ and product scores $\{u_{\alpha}^{(n)}\}$ in a recursive way
\begin{equation}
\begin{split}
d_{i}^{(n)}&=\frac{1}{d_{i}}\sum_{\alpha}M_{i\alpha}\,u_{\alpha}^{(n-1)},\\
u_{\alpha}^{(n)}&=\frac{1}{u_{\alpha}}\sum_{i}M_{i\alpha}\,d_{i}^{(n-1)},
\end{split}
\label{mr}
\end{equation}
where $M$ is the adjacency matrix of the country-product network and $d_{i}^{(0)}=d_{i}$, $u_{\alpha}^{(0)}=u_{\alpha}$. 
Ref. \cite{hidalgo2009building} considers $d^{(n)}$ and $u^{(n)}$ as generalized measures of diversification and ubiquity, respectively.
Products are ranked in order of increasing $u_{\alpha}^{(2n)}$ consistently with the economic interpretation: complex products tend to be less ubiquitous.
The Economic Complexity Index based on $d^{(n)}$ and $u^{(n)}$ is a better predictor of the future economic growth of 
a country than the existing indicators that do not take the network connectedness into account, such as institutional and education quality 
measures \cite{hausmann2014atlas}.
However, this method presents both mathematical and conceptual
issues \cite{tacchella2012new,cristelli2013measuring,caldarelli2012network,tacchella2013economic,battiston2014metrics}.
In particular, the interpretation of the scores changes when considering odd or even iteration order $n$, high-order iterations are difficult to interpret,  
and the process asymptotically converges to a trivial fixed point
\cite{cristelli2013measuring,caldarelli2012network}.
The problem of convergence to an uniform fixed point can be bypassed by 
defining the final country score $E_{i}$ as
$E_{i}=(d_{i}^{(2n)}-\overline{d^{(2n)}})/\sigma_{d^{(2n)}}$, where $\overline{d^{(2n)}}$ and $\sigma_{d^{(2n)}}$ are the average and the standard deviation
of scores $d_{i}^{(2n)}$, respectively.
For $n$ sufficiently large, $E_{i}$ is proportional to the $i$-th component of the eigenvector associated to the second largest eigenvalue of the 
stochastic matrix relating $d^{(2n+2)}$ with $d^{(2n)}$ \cite{caldarelli2012network}.
When the number of iterations becomes large, score differences between countries can become smaller than the computational precision,
which is why Ref. \cite{hidalgo2009building} considers only MR variables
for $n\leq19$.
We consider here only the even iterations and, given that the process described by Eq. \eqref{mr} converges to a uniform fixed point, 
we study the ranking for different values of $n$
in the range $n\in[0,20]$.

\subsection{Fitness-Complexity Method (FCM) and its generalization}

In the original Fitness-Complexity Method (FCM),
the country fitness $\{F_{i}\}$ and product complexity values $\{Q_{\alpha}\}$ are 
defined as the stationary state of the following non-linear iterative process \cite{tacchella2012new}
\begin{equation}
\begin{split}
  \tilde{F}_{i}^{(n)}&=\sum_{\alpha}M_{i\alpha}Q_{\alpha}^{(n-1)},\\
  \tilde{Q}_{\alpha}^{(n)}&=\frac{1}{\sum_{i}M_{i\alpha}\frac{1}{F^{(n-1)}_{i}}}
\end{split}
\label{metrics}
\end{equation}
where scores are normalized after each step according to
\begin{equation}
\begin{split}
  F_{i}^{(n)}=\tilde{F}_{i}^{(n)}/\,\overline{F^{(n)}},\\
  Q_{\alpha}^{(n)}=\tilde{Q}_{\alpha}^{(n)}/\,\overline{Q^{(n)}},
\end{split}
\end{equation}
with the initial condition $F_{i}^{(0)}=1$ and $Q_{\alpha}^{(0)}=1$.
The metric defined by Eq. \eqref{metrics} has been shown to be economically well-grounded \cite{tacchella2012new,cristelli2013measuring} and
to be highly informative about the future economic development of countries \cite{cristelli2015heterogeneous}.
Moreover, the metric has been recently applied beyond its original scope: it has been shown to be 
the most efficient metric among several network-based metrics in ranking species according to their importance 
in mutualistic ecological networks \cite{dominguez2015ranking}. The method has also provided new insights
into the long-lasting problem of evaluating the scientific competitiveness of nations \cite{cimini2014scientific}.

In general, the parameter dependence of an algorithm is an important problem. For example, the dependence of the prominent PageRank
algorithm on its only parameter, the teleportation parameter, has been studied in detail (see \cite{berkhin2005survey} for a review).
We study here the generalized FCM defined by the equations \cite{pugliese2014convergence}
\begin{equation}
\begin{split}
  \tilde{F}_{i}^{(n)}(\gamma)&=\sum_{\alpha}M_{i\alpha}Q_{\alpha}^{(n-1)},\\
  \tilde{Q}_{\alpha}^{(n)}(\gamma)&=\Biggl[\sum_{i}M_{i\alpha}\bigl(F^{(n-1)}_{i}\bigr)^{-\gamma}\Biggr]^{-1/\gamma}.
\end{split}
\label{newmetrics}
\end{equation}
We refer to the scores produced by these equations as generalized fitness $F(\gamma)$ and complexity $Q(\gamma)$ and to $\gamma$ as the extremality parameter.
Eq. \eqref{newmetrics} reduces to Eq. \eqref{metrics} for $\gamma=1$ ($F(1)=F$).
Consistently with the economic interpretation of the equations, we restrict our analysis to $\gamma>0$ \cite{pugliese2014convergence}.
The higher the value of $\gamma$, the more sensitive the generalized complexity is to the fitness of the least-fit exporting country.
When $\gamma$ is sufficiently small we observe a transition to
a condensed phase
where all score is accumulated by the least ubiquitous product and the country exporting it. To avoid this transition,
we study only $\gamma>0.6$ in the following.

The convergence properties of the algorithm defined by Eq. \eqref{newmetrics} are highly non-trivial due to the non-linear coupling
between fitness and complexity.
We do not attempt to study how the convergence properties of the algorithm depend on $\gamma$ and
simply run $I=1000$ iterations of the process defined by Eq. \eqref{newmetrics}.
In our dataset $I=1000$ is a reasonable choice, because ranking switches
are rare for $I>1000$ (in agreement with the results presented in Ref. \cite{pugliese2014convergence}), involve only
low-ranked countries (or products) and their effect on our results is negligible.
We also notice that after a larger number of iterations some scores are rounded to zero due to machine precision, which causes a 
loss of the method's discriminative power.
Our choice $I=1000$ allows us to avoid the appearance of these zero scores and the consequent loss of discriminative power.

\begin{figure*} [t]
\centering
  \includegraphics[height=0.85\columnwidth, angle=270]{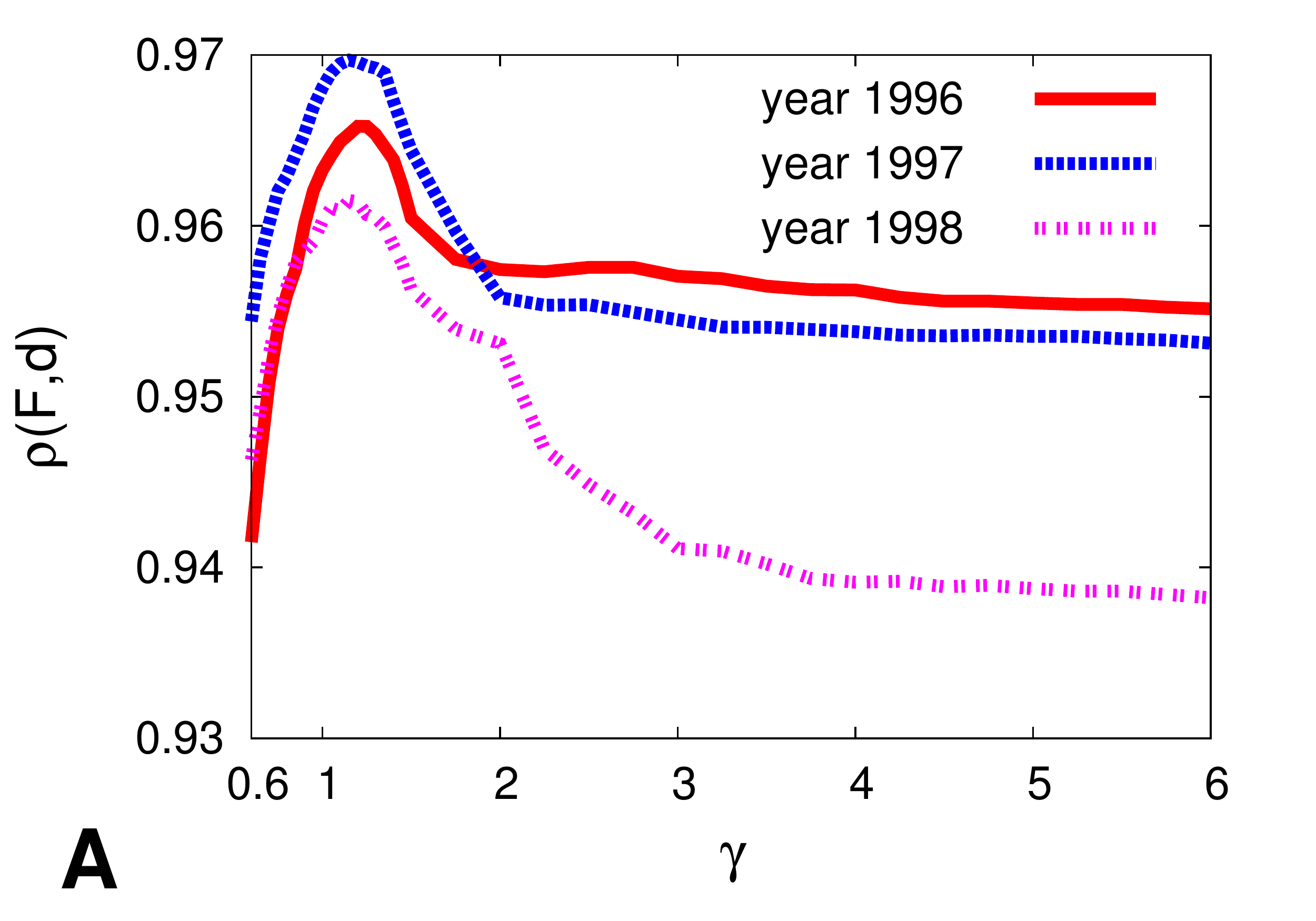}\includegraphics[height=0.85\columnwidth, angle=270]{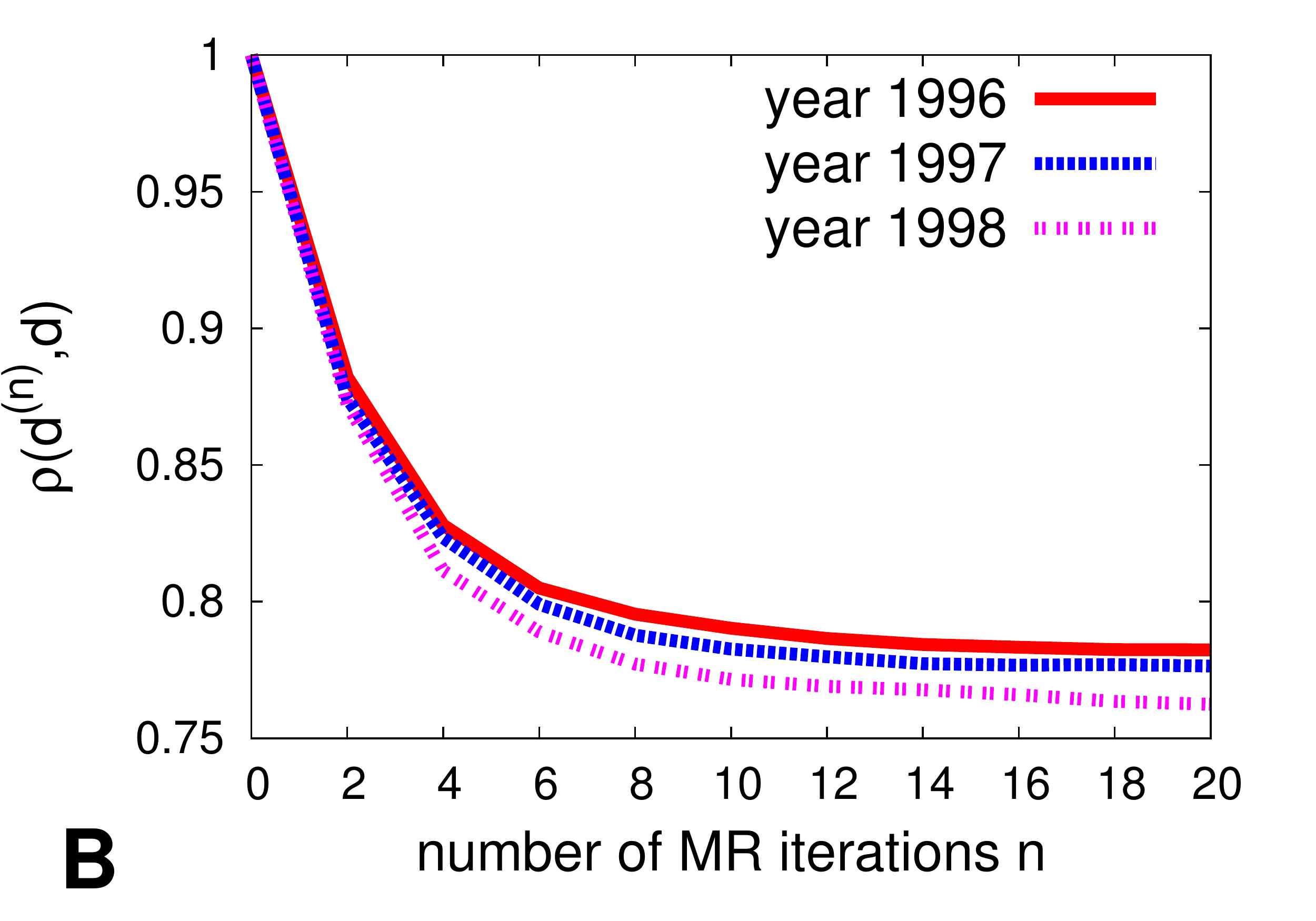}
\caption{Spearman's correlation $\rho$ between country score and diversification.
\emph{Panel A:} $\rho(F(\gamma)),d)$ as a function of $\gamma$ (FCM). \emph{Panel B:} $\rho(d^{(n)},d)$ as a function of the number of MR iterations $n$.}
\label{fig:div}       
\end{figure*}

\begin{figure*} [t]
 \centering
  \includegraphics[height=0.85\columnwidth, angle=270]{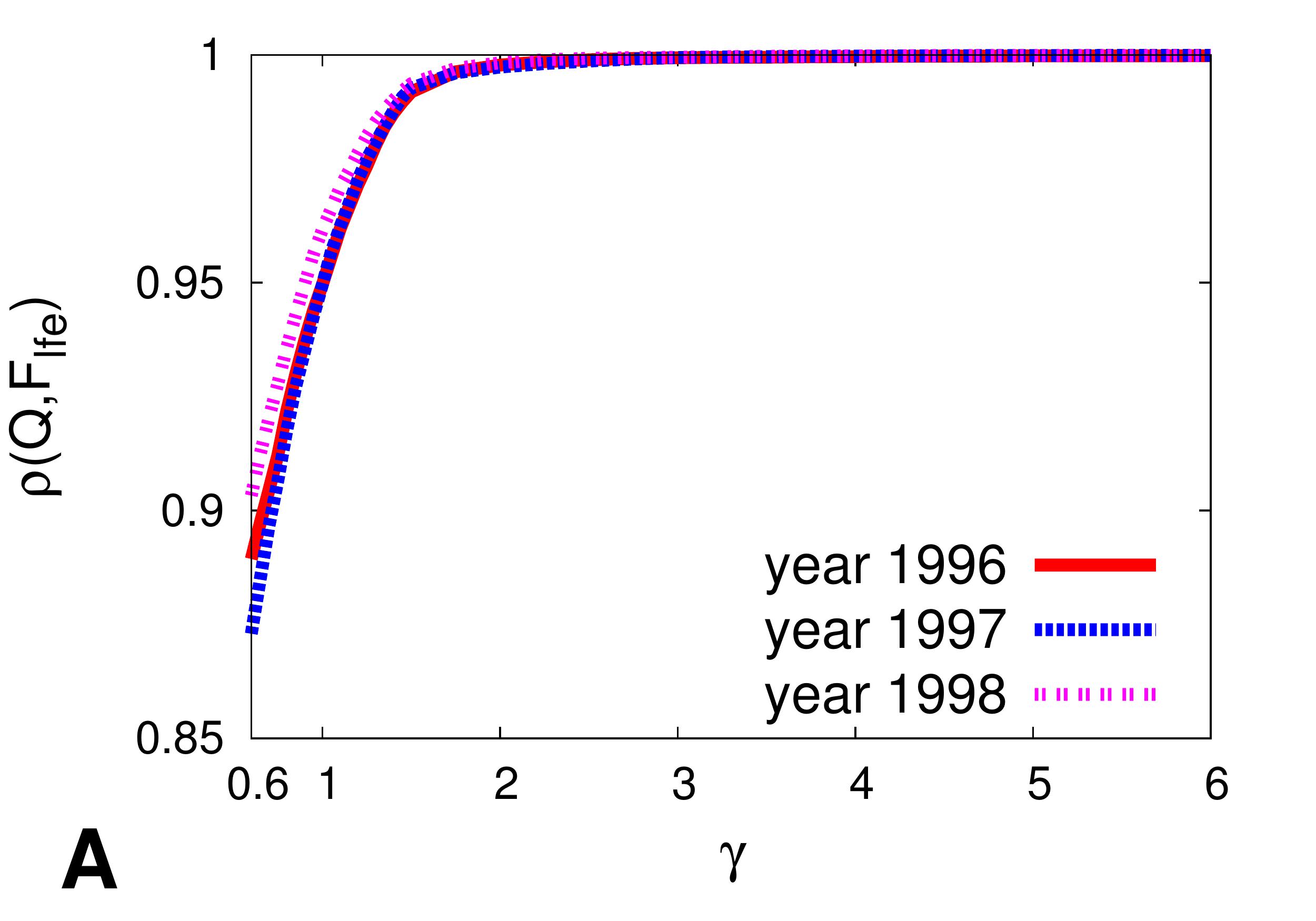}\includegraphics[height=0.85\columnwidth, angle=270]{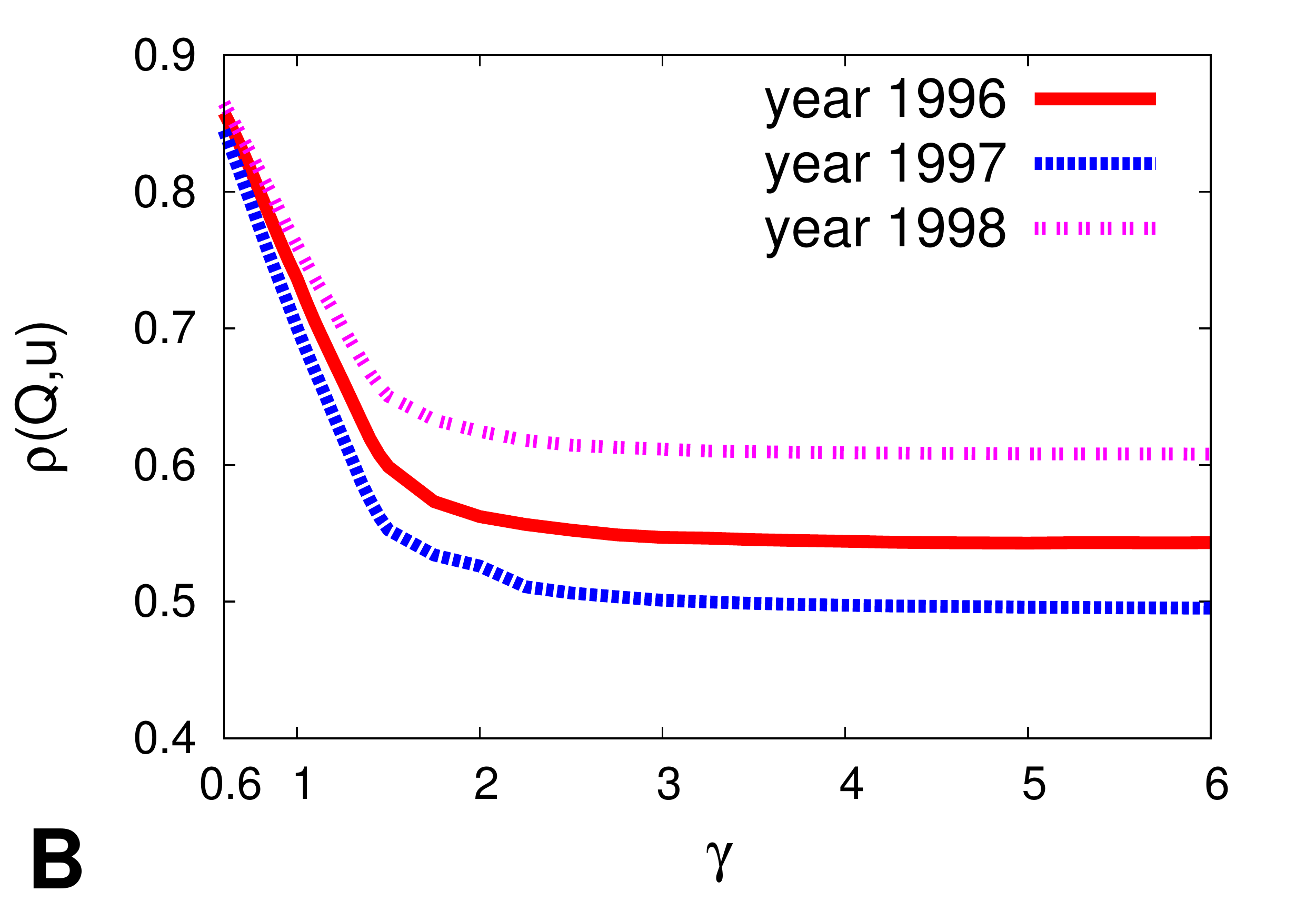}
\caption{The influence of $\gamma$ on the product ranking by the FCM.
\emph{Panel A:}
Spearman's correlation between product complexity $Q(\gamma)$ and the ranking of product according
to the fitness $F_{lfe}^{(\gamma)}$ of the least-fit exporting countries.
\emph{Panel B:} Spearman's correlation between complexity $Q(\gamma)$ and ubiquity $u$.}
\label{fig:gamma}       
\end{figure*}

\begin{figure*} [t]
\centering
  \includegraphics[height=0.85\columnwidth, angle=270]{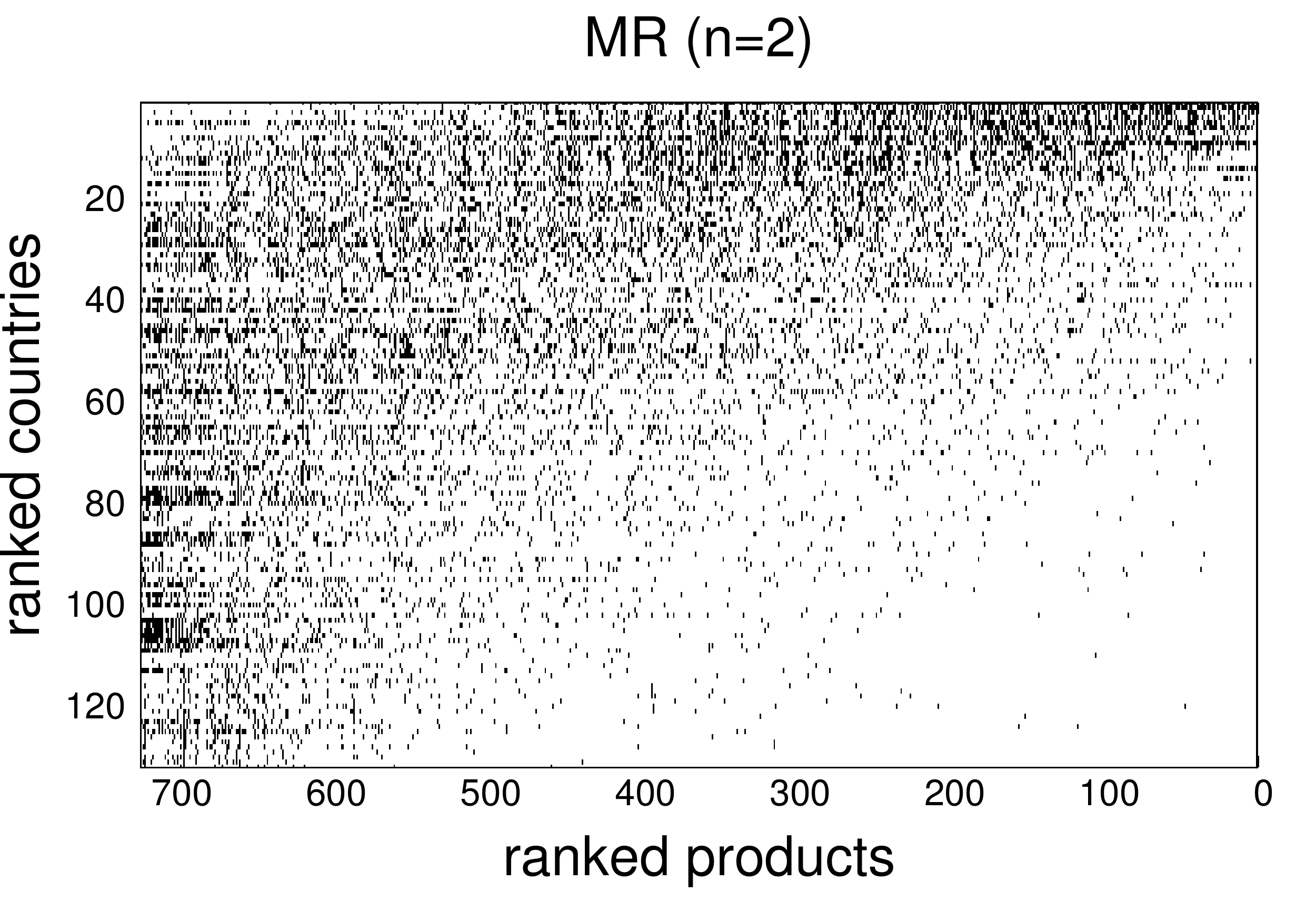}\includegraphics[height=0.85\columnwidth, angle=270]{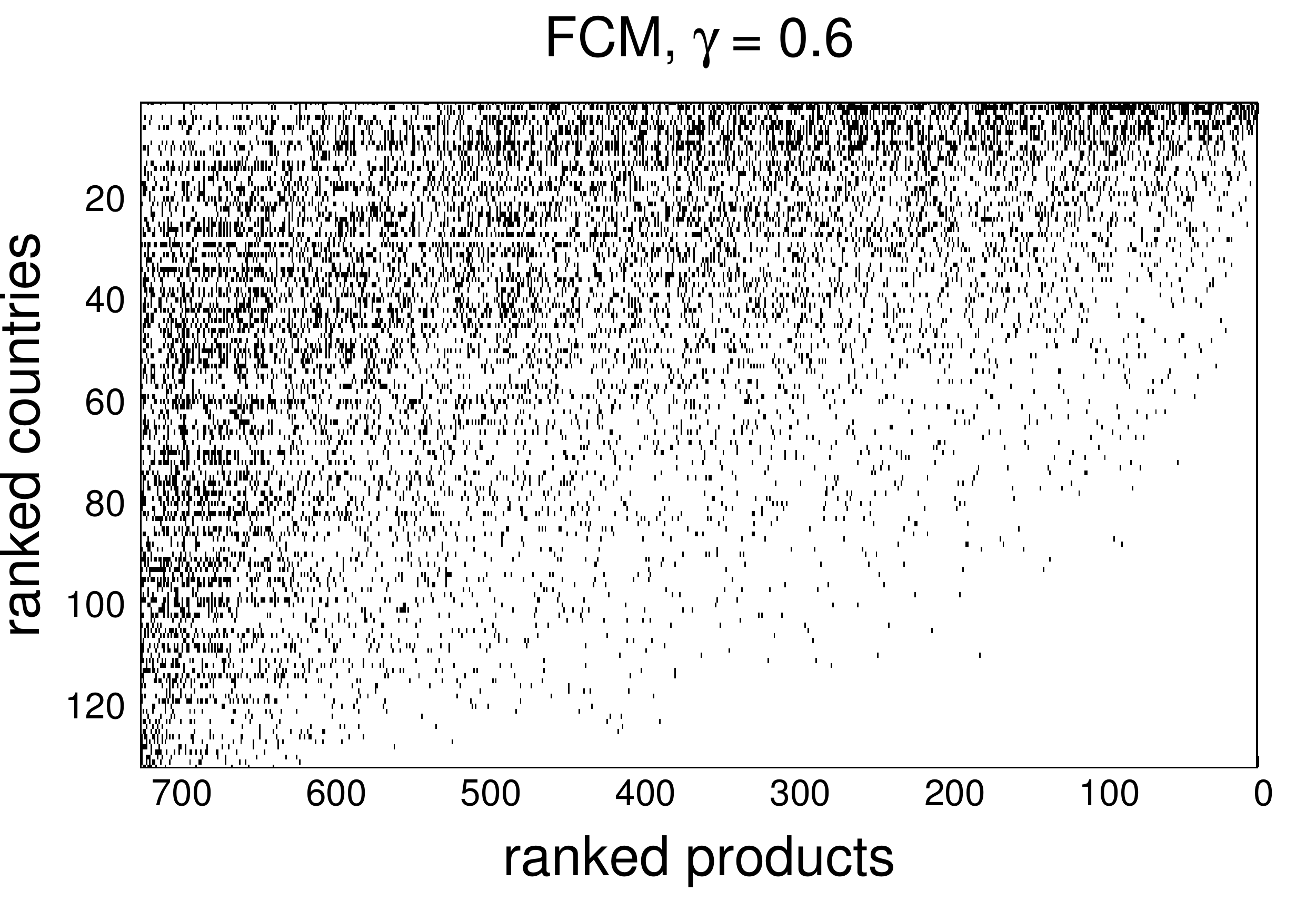}
  \includegraphics[height=0.85\columnwidth, angle=270]{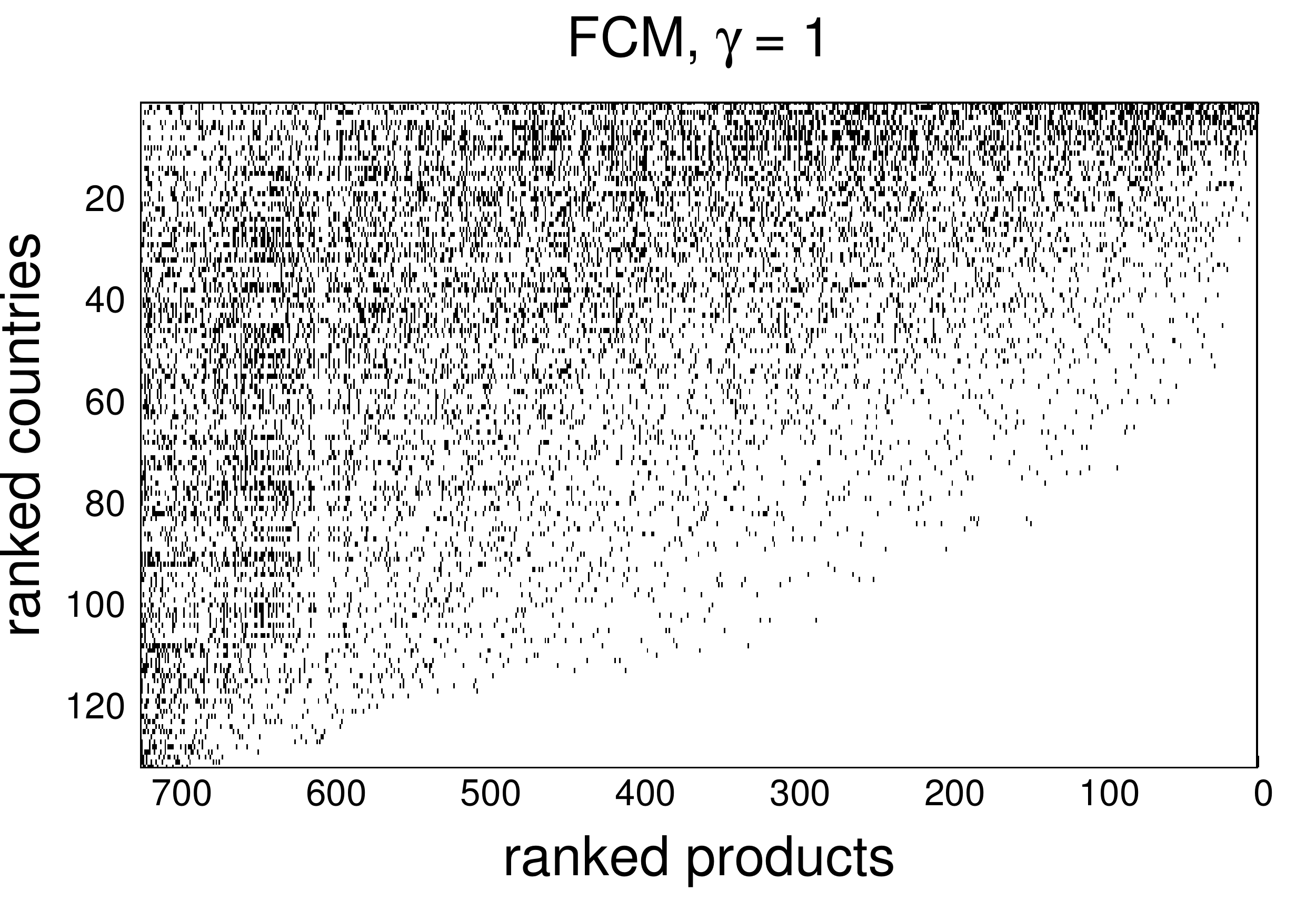}\includegraphics[height=0.85\columnwidth, angle=270]{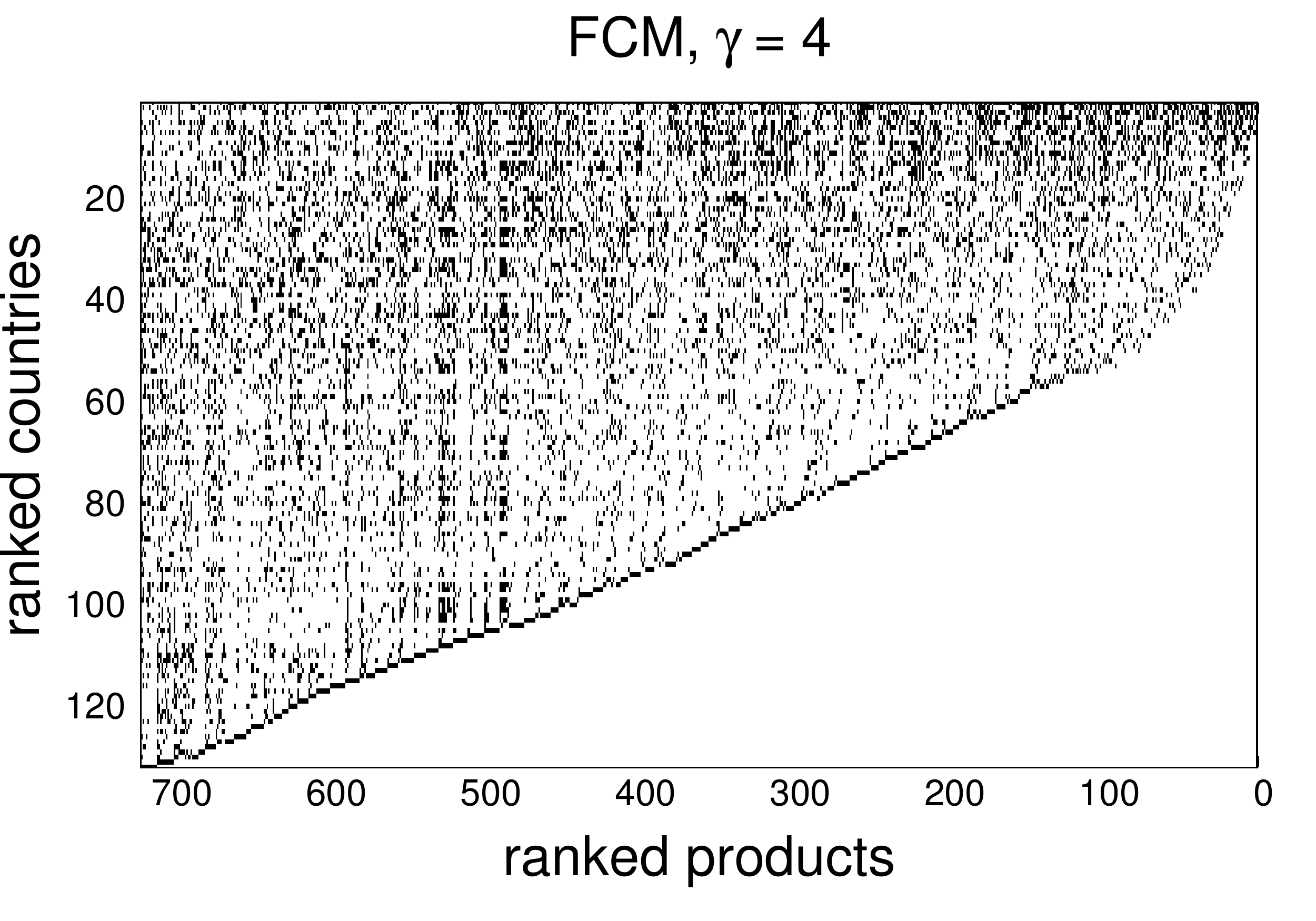}
\caption{Country-product matrices resulting from different metrics (1996).
The FCM produces a more triangular country-product matrix than the MR.
The matrix produced by the MR after $2$ iterations has many points lying close to the bottom-right corner of the matrix
which means that some products are ranked high yet produced by low-ranked countries.
Further iterations of the MR do not improve the triangularity of the matrix.
For the FCM, triangularity improves as $\gamma$ increases. Note that the
border between
the filled and the empty region of the matrix
is sharper for $\gamma=4$ than for $\gamma=1$. This is due to the extremality of the ranking:
for $\gamma=4$, products are (almost) perfectly ranked according to the 
fitness of their least-fit exporter (see Fig. \ref{fig:gamma}).}
\label{fig:matrices}       
\end{figure*}

\begin{figure*} [t]
\centering
  \includegraphics[height=0.85\columnwidth, angle=270]{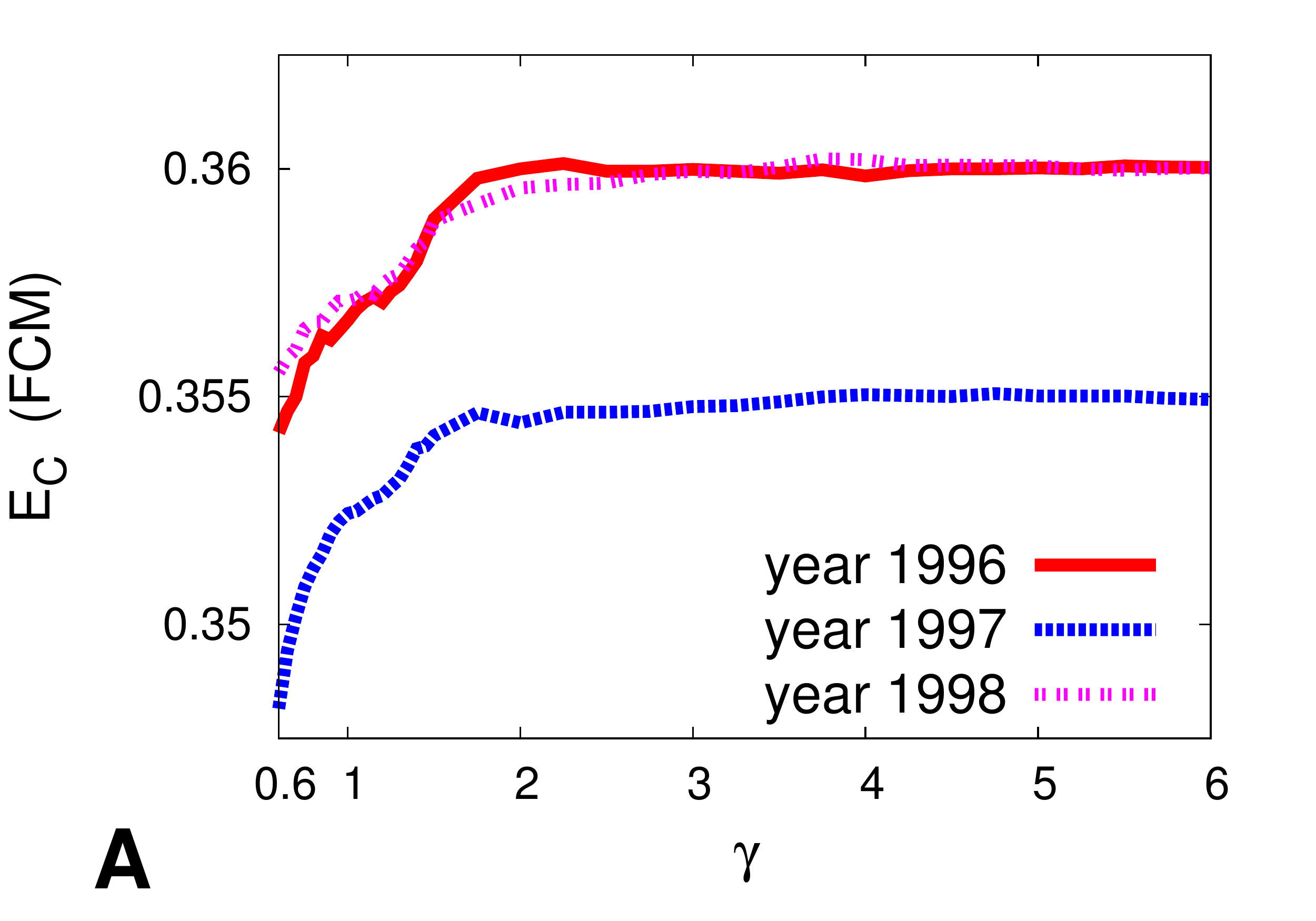}\includegraphics[height=0.85\columnwidth, angle=270]{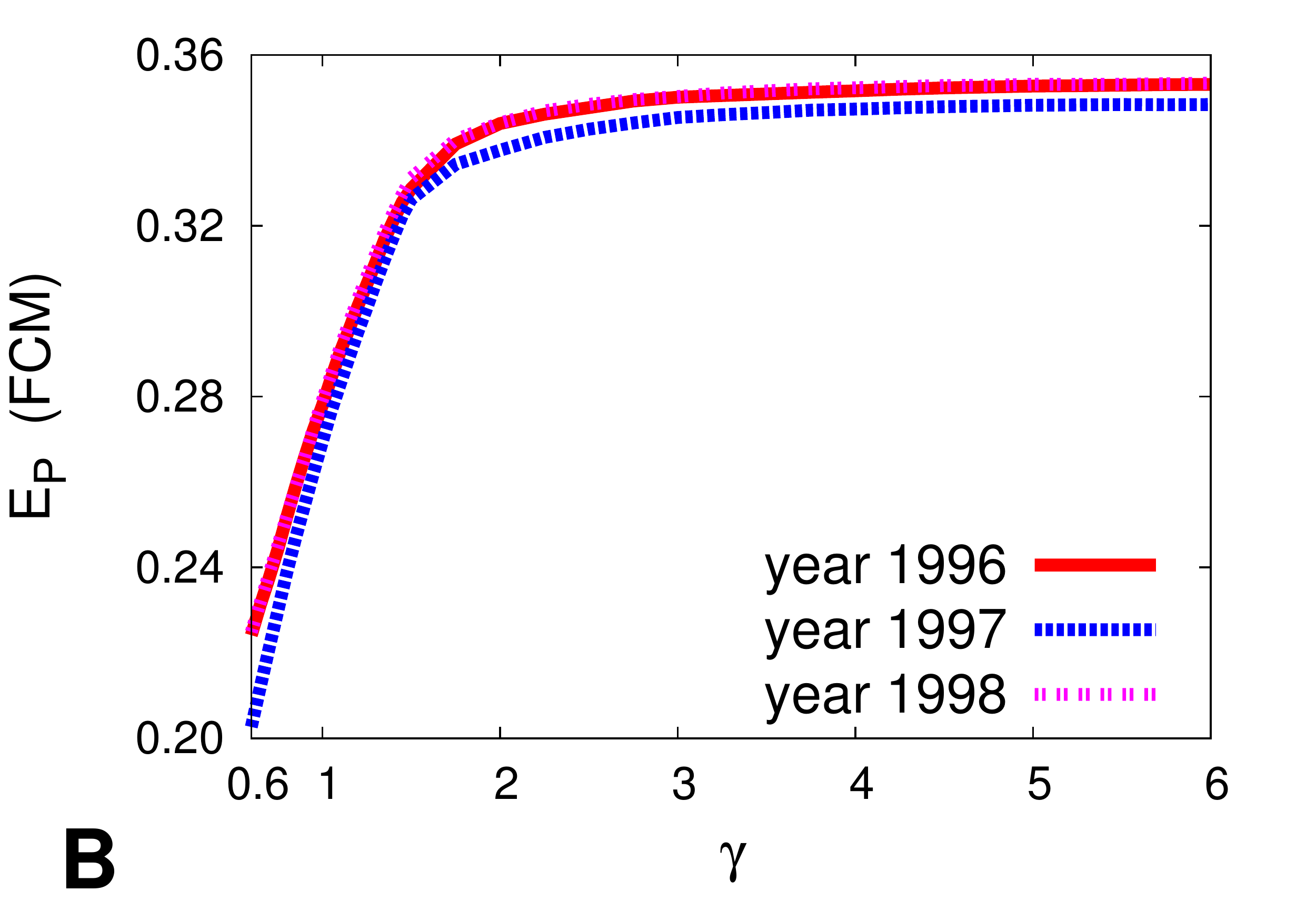}
  \includegraphics[height=0.85\columnwidth, angle=270]{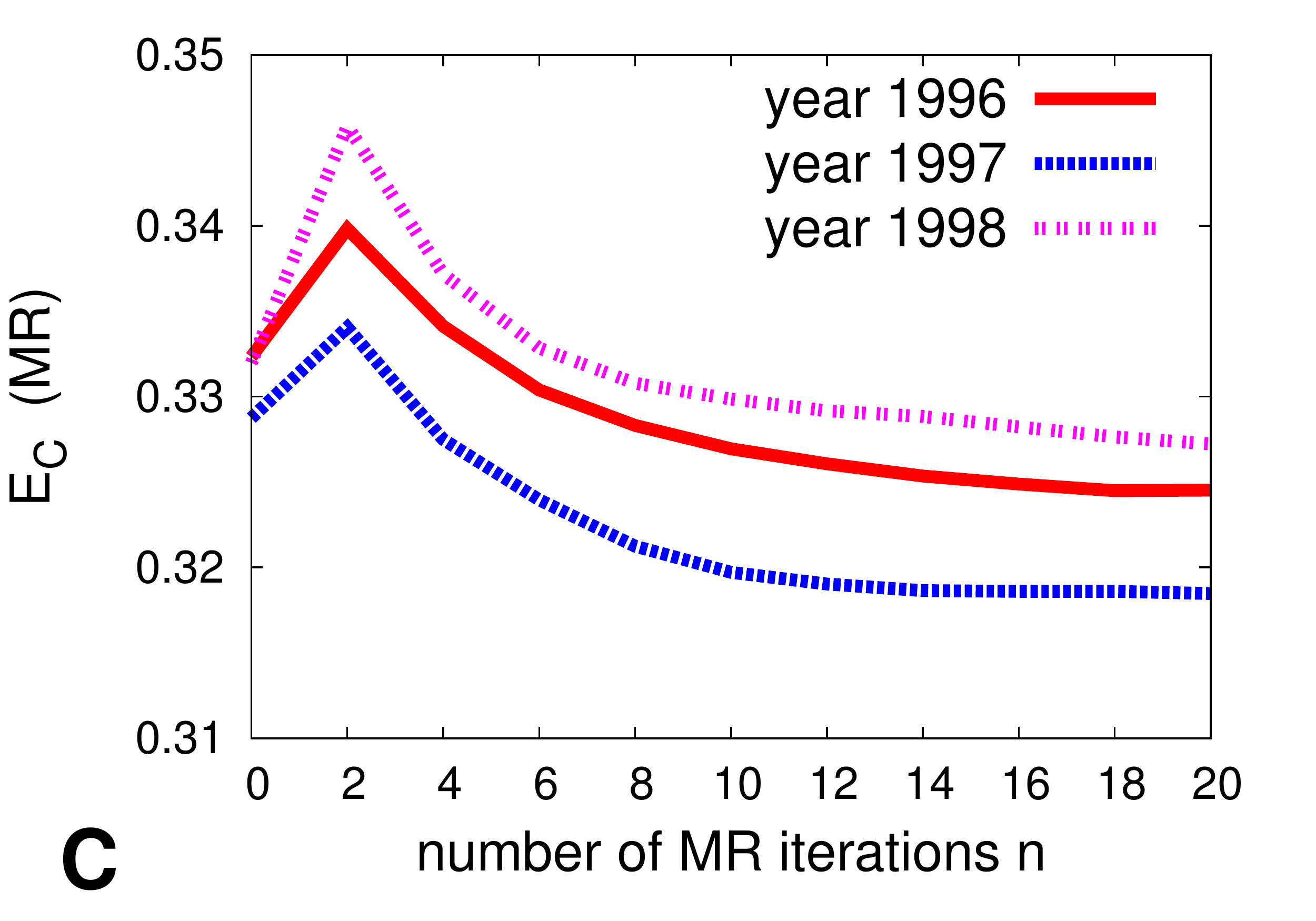}\includegraphics[height=0.85\columnwidth, angle=270]{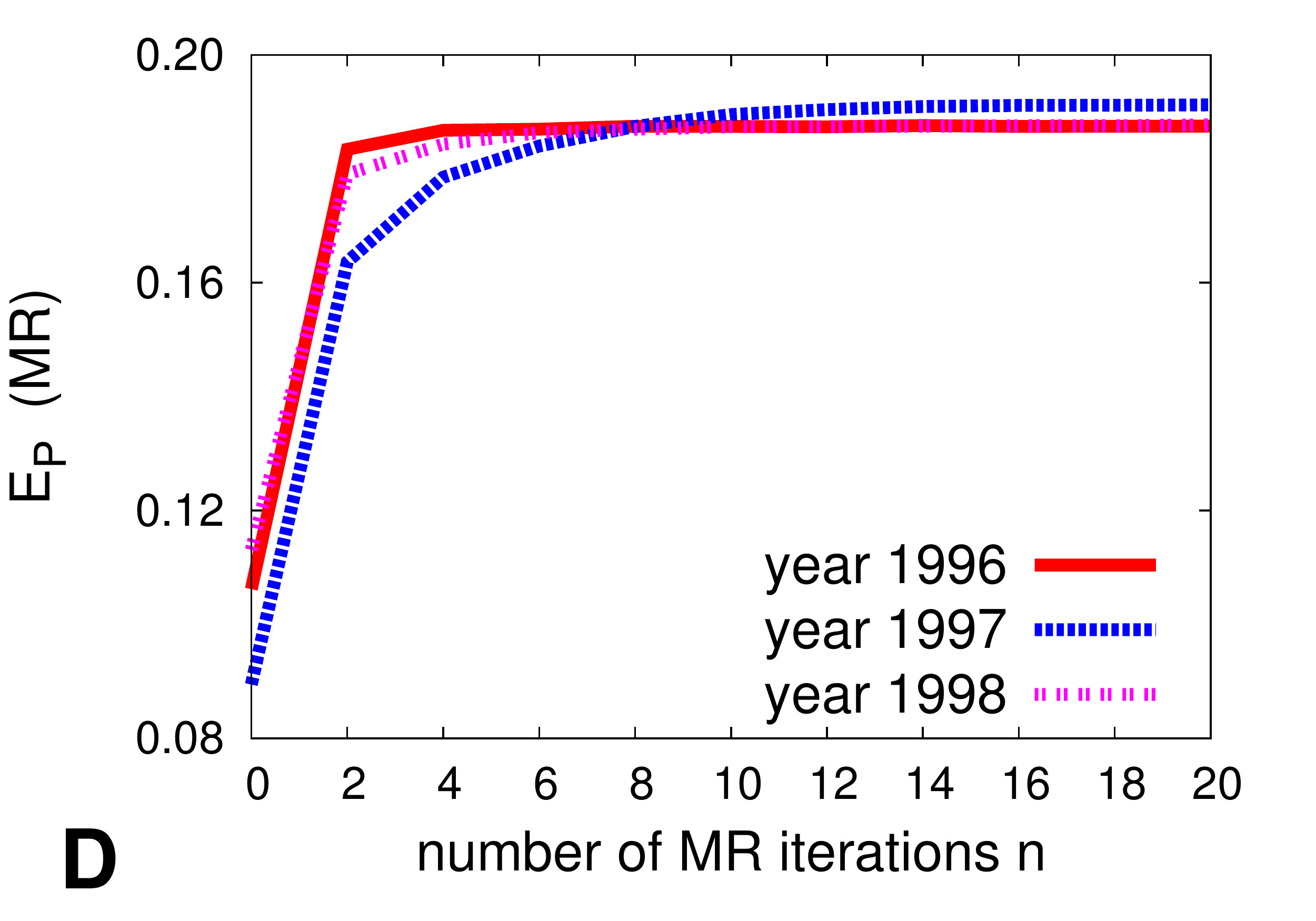}
\caption{Extinction areas for countries and products for the FCM and the MR.
Panels A and B show the extinction areas for countries and products, respectively, for the FCM
as a function of the extremality parameter $\gamma$.
Panels C and D show the extinction areas for countries and products, respectively, for the MR as a function of the number of iterations $n$.
}
\label{fig:extinction}       
\end{figure*}

\begin{figure*} [t]
\centering
  \includegraphics[height=1\columnwidth, angle=270]{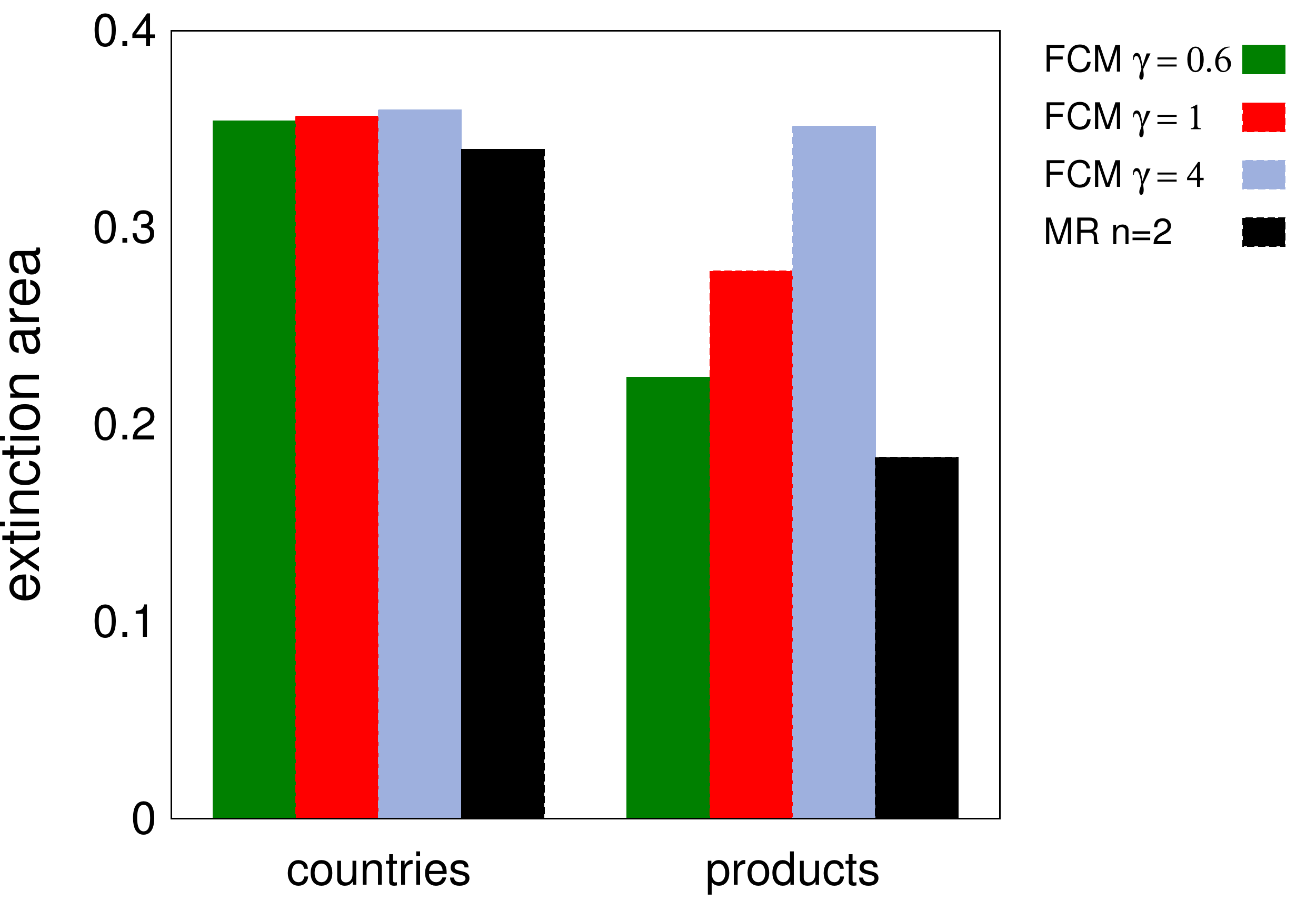}
\caption{Ability of algorithms to rank the nodes by importance (year 1996). The original FCM ($\gamma=1$, red bar) 
outperforms the MR according to the extinction areas
for countries and products. The extinction areas for the generalized FCM with $\gamma>1$ and $\gamma<1$ are larger and smaller, 
respectively, than the extinction areas
for the original FCM $\gamma=1$.}
\label{fig:algorithm_comparison}       
\end{figure*}

\begin{figure*} [t]
\centering
  \includegraphics[height=0.85\columnwidth, angle=270]{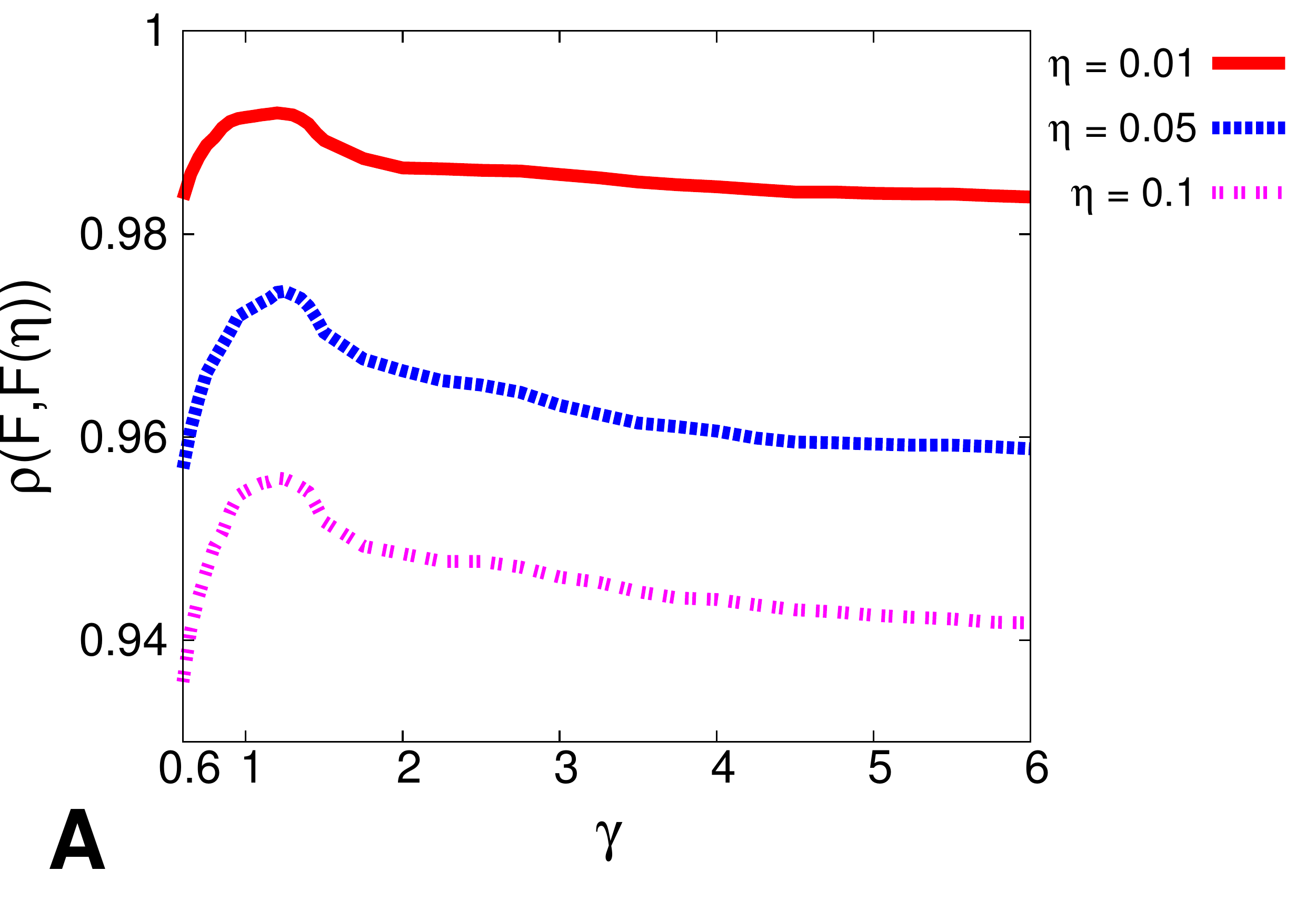}\includegraphics[height=0.85\columnwidth, angle=270]{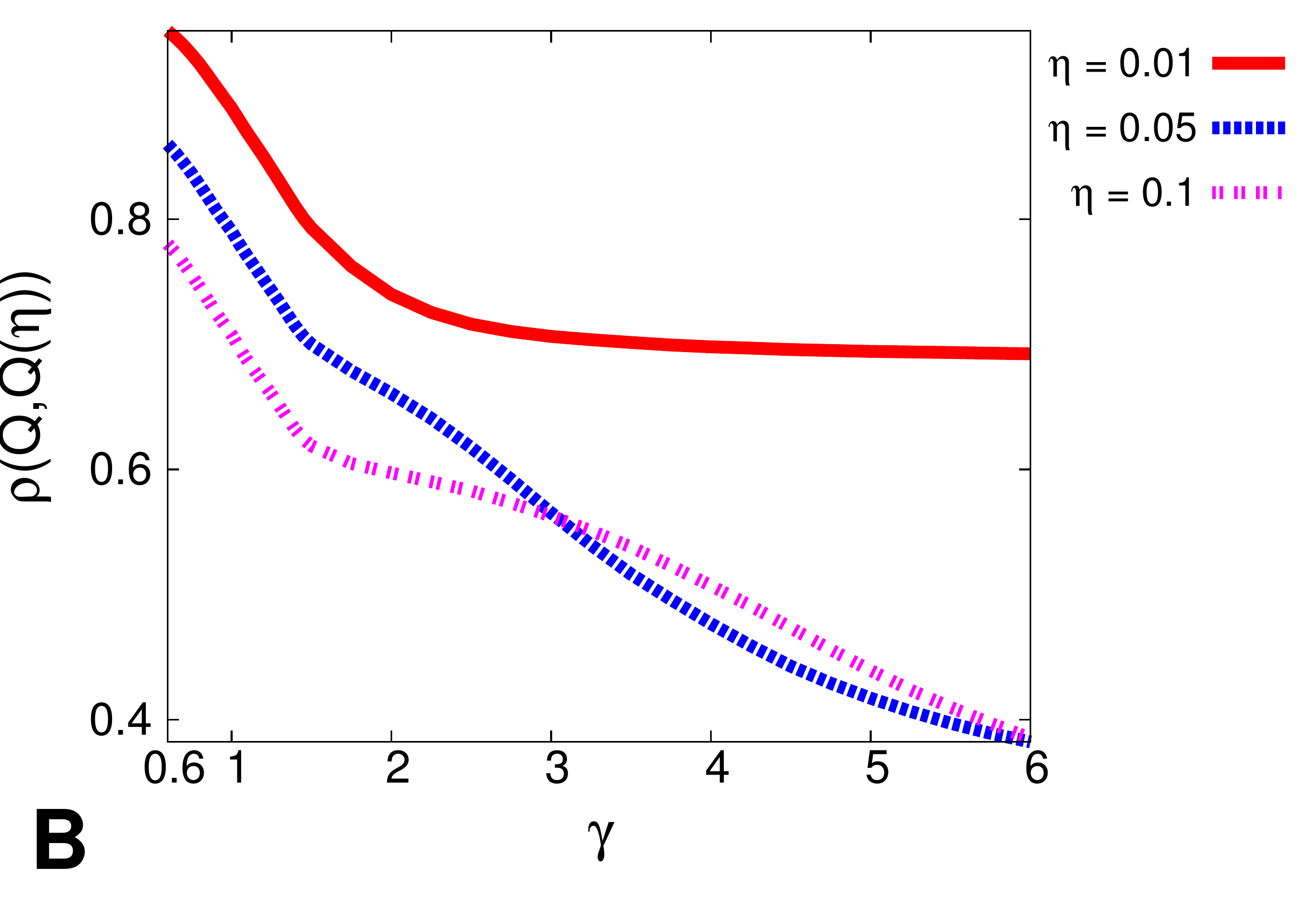}
\caption{Robustness against noise of the rankings by the FCM as a function of the extremality parameter $\gamma$. 
Robustness is measured by the Spearman's correlation
between the rankings before and after the inversion of a fraction $\eta$ of random bits in the country-product matrix (year 1996).
Panels A and B show results for countries and products, respectively.}
\label{fig:stability}       
\end{figure*}

\begin{figure*} [t]
\centering
  \includegraphics[height=0.85\columnwidth, angle=270]{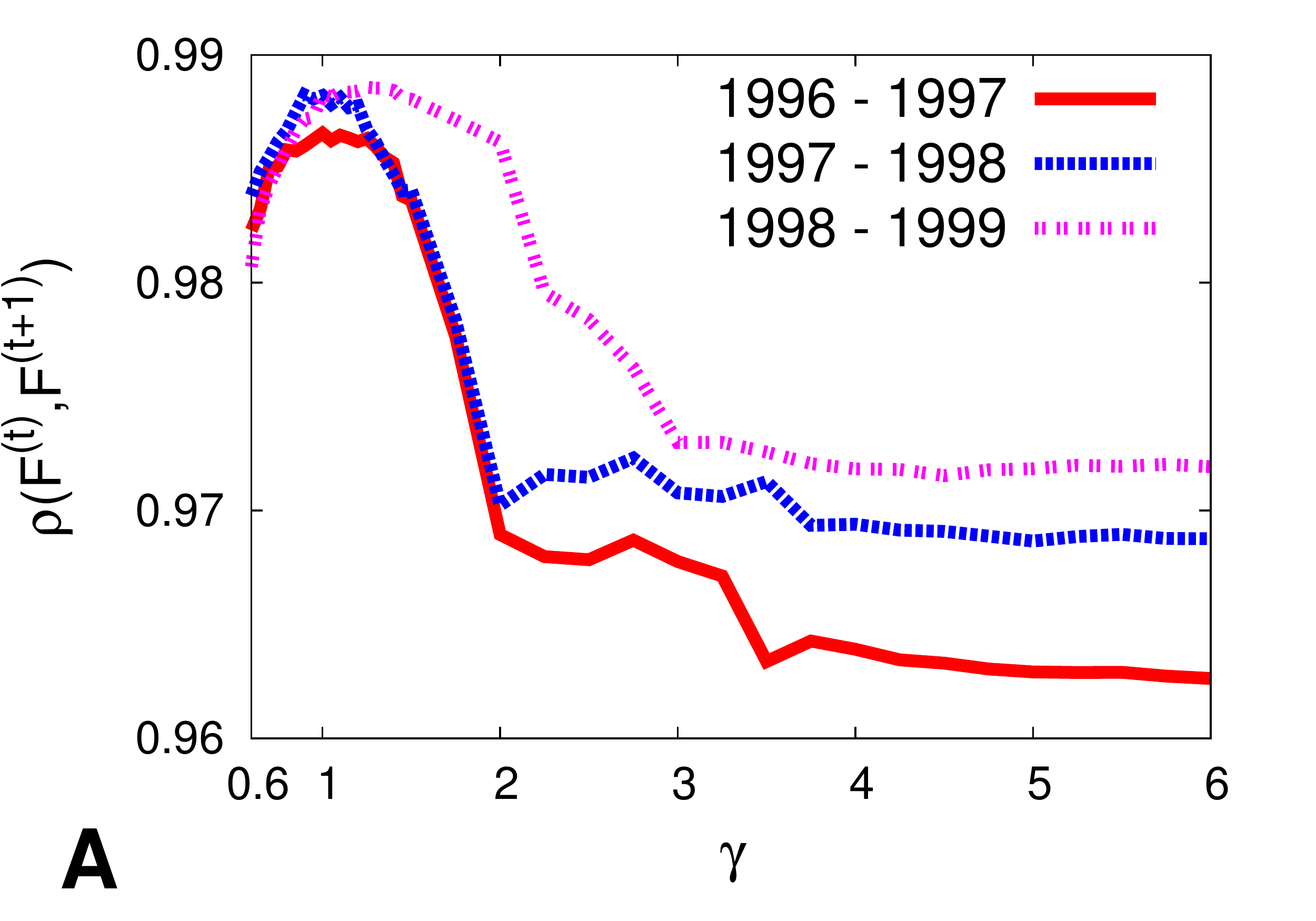}\includegraphics[height=0.85\columnwidth, angle=270]{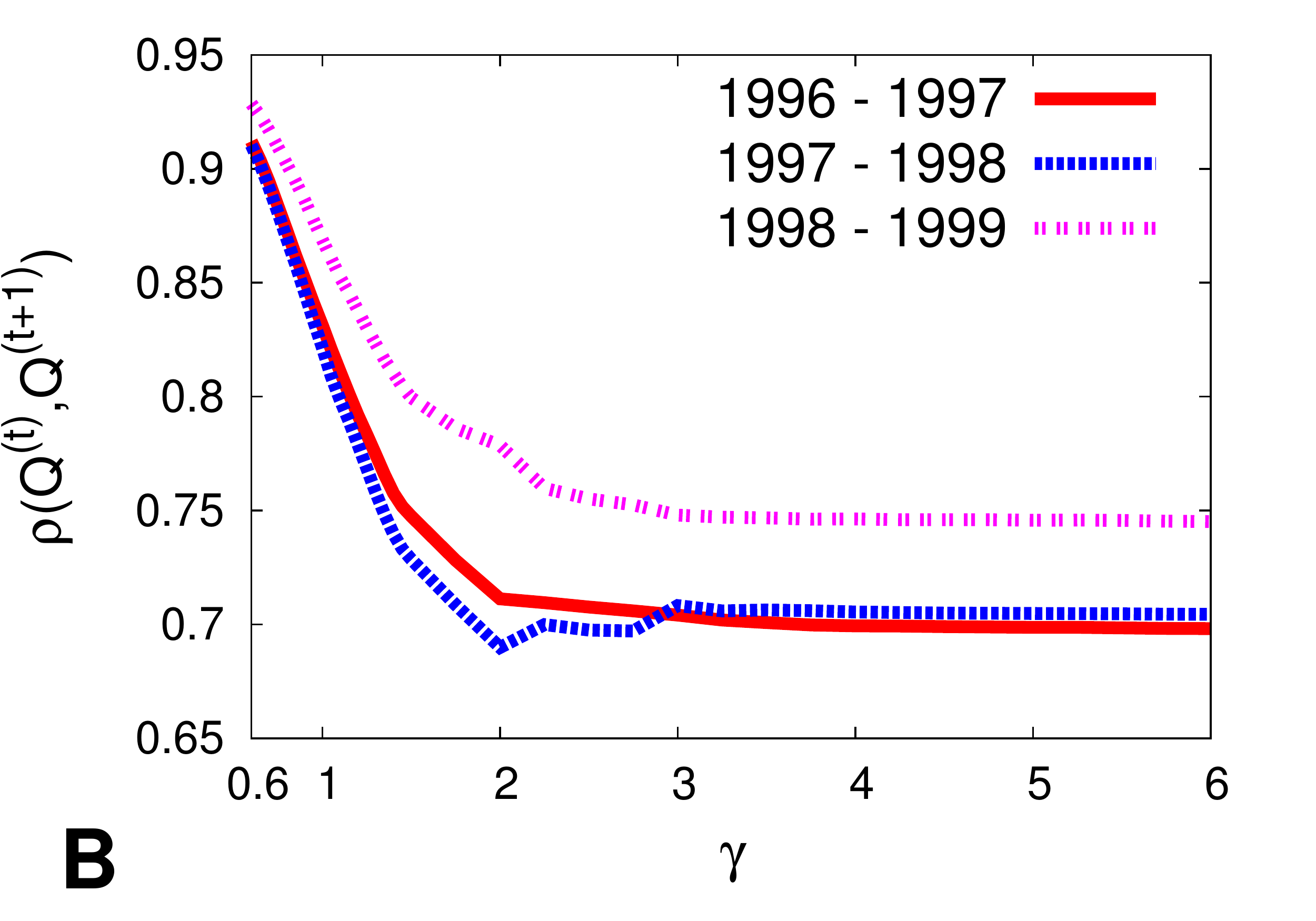}
\caption{Ranking volatility as a function of the extremality parameter $\gamma$ for the FCM.
Volatility is measured by the Spearman's correlation between the rankings in two consecutive years. 
Panels A and B show results for countries and products, respectively.}
\label{fig:volatility}       
\end{figure*}

\section{Results}

Our main aim is to compare the metrics for economic complexity with respect to different criteria.
Here we discuss the results on the NBER-UN dataset described in the Materials and Methods section, focusing on three different years (1996-1998);
the results for the different years are in qualitative agreement.

\subsection{Correlation between complexity of countries and diversification}

According to the economic complexity interpretation of the international trade, 
the level of diversification of a country's exports reflects its
level of industrial development.
Indeed, in order to produce a certain product, a country must own a variety of appropriate capabilities, which may include suitable climatic conditions, 
labor skills, strong scientific research,
and others.
While these capabilities are not measurable in the real world, we can infer a country's complexity by looking at its export basket:
a developed country owns a large number of capabilities and, as a result, is able to export many products.
Consequently, we expect a good metric to produce a country ranking that is highly correlated with the 
ranking by diversification $d$, i.e., with the ranking by number
of exported products.
Fig. \ref{fig:div} shows the Spearman's correlation between the rankings produced by different metrics and diversification.
For the FCM, $\rho(F(\gamma),d)$ is maximal when $\gamma\simeq1.2$ and decreases as $\gamma$ becomes significantly larger than one.
For the MR, $\rho(d^{(0)},d)=1$ by definition, and $\rho(d^{(n)},d)$ decreases when the number of MR iterations increases, which means
that the information about diversification is progressively lost. After two iterations, the 
correlation of the MR country score with diversification is already considerably
smaller than
the correlation between fitness and diversification.
This loss of information caused by MR iterations is qualitatively in agreement with results on artificial data based on a toy model
where capabilities are explicitly defined
(see Fig. 12 of Ref. \cite{cristelli2013measuring}). The lower correlation with diversification 
is a negative issue for the MR in view 
of the interpretation of country complexity as a measure of the diversity of its available capabilities.

\subsection{Role of the extremality parameter $\gamma$}

We compare now the product rankings produced by the FCM for different values of $\gamma$.
When $\gamma$
is large, the sum $\sum_{i}M_{i\alpha}\,F_{i}^{-\gamma}$ in Eq. \eqref{newmetrics} becomes dominated by 
the term corresponding to the least-fit exporter country, 
while all other terms become negligible. As a result, when $\gamma\gtrsim 2$ the ranking of countries becomes almost perfectly extremal, i.e., almost
perfectly correlated with the ranking of products according to the fitness values of their least-fit exporting countries, as shown in Fig. \ref{fig:gamma}A.
Conversely, when $\gamma$ is close to one, $Q(\gamma)$ is also sensitive to the number of terms
involved in the sum of $F^{-\gamma}$ terms, i.e., to product ubiquity.
For this reason, the Spearman's correlation $\rho(Q(\gamma),u)$ between product complexity $Q(\gamma)$ and ubiquity monotonously decreases with $\gamma$,
as shown in Fig. \ref{fig:gamma}B.

\subsection{Evaluating economic complexity rankings}

As mentioned before, competitive countries have a highly diversified export basket, while developing countries tend to export ubiquitous products. 
This property is well captured by the Fitness-Complexity metric \cite{tacchella2012new}: when 
arranging the country-product matrix according to ranking, the matrix
has a triangular shape.
This is also a feature found in ecological systems, such as plant-animal mutualistic networks, where it is 
referred to as nestedness \cite{bascompte2003nested,bascompte2010structure}.
Fig. \ref{fig:matrices} shows the country-product matrices generated by different metrics. Note that the FCM produces a more triangular shape than the MR,
meaning that the FCM better captures the nested structure of the international trade than the MR.
Moreover, the triangularity improves when $\gamma$ increases.

To assess the quality of economic complexity algorithms, 
we use the extinction area metric which has been introduced in the context of ecological networks.
In an ecological network where active species are connected with passive species,
the extinction area quantifies the ability of a certain algorithm to rank active (or passive) species by importance (or vulnerability)
when active species (or passive) are
progressively removed from the system in order of decreasing importance (or increasing vulnerability)
\cite{allesina2009googling,dominguez2015ranking}.
According to this metrics, a good active-species ranking leads to a quick breakdown of the system when active species are
progressively removed from the system in order of decreasing importance.
We use this metric to evaluate algorithms for economic complexity, assuming that the most important countries and vulnerable products 
are the most complex
countries and products, respectively.
This assumption is motivated by the observation that both kinds of systems, ecological networks and the country-product matrix,
have a nested structure.
For the country-product network, the nested structure implies that
the artificial deletion of a complex country must have a strong impact on the system, because it
is likely to cause the disappearance of the most complex products from the global market. Conversely,
the presence of developing countries tends to be less essential for the existence of products in the market.
Similarly, removing non-complex products must strongly impact the economy of developing countries, 
which are likely to disappear when a sufficient number of non-complex products
are removed from the system. In contrast, competitive countries can survive in the market even if many low-quality products are removed from the trade,
because of their diversified export basket.

Based on this intuition, a good metric for economic complexity is expected to maximize 
the extinction areas for countries (hereafter $E_C$) and for products (hereafter $E_P$).
To define $E_C$, countries (with all their links) 
are sequentially removed from the network in order of decreasing ranking, from the first in the ranking to the last one.
This procedure defines an extinction curve $e_{C}(f_{C})$ for countries, which is the fraction of extinct products as a function of the fraction 
$f_C$ of removed countries;
$E_C$ is defined as the area below the curve $e_{C}(f_{C})$ from $f_{C}=0$ (no removed countries) to $f_C=1$ (all countries have been removed).
This area is large when important countries are ranked high and are consequently removed early.
The extinction area for products $E_P$ is defined analogously as the area below the curve $e_{P}(f_P)$ from $f_P=0$ to $f_P=1$, where $e_P(f_P)$
is the fraction of extinct countries after the removal of a fraction $f_P$ of products (products are removed in order of increasing ranking).
Following Ref. \cite{dominguez2015ranking},
the rankings are not recomputed after every node removal.
In the case of ranking ties we consider $100$ different randomized rankings and compute the average extinction area.

Fig. \ref{fig:extinction} (panels A-B) shows the dependence of $E_C$ and $E_P$ on the parameter $\gamma$ for the generalized FCM:
the extinction areas increase with $\gamma$ when $\gamma\in [0.6,2]$, ending in a plateau for $\gamma>2$, approximately.
Fig. \ref{fig:extinction} (panels C-D) shows the dependence of $E_C$ and $E_P$ on the number of iterations $n$ for the MR.
For the country ranking, the first two MR iterations (the diversification $d_i=d^{(0)}_i$ of 
country $i$ and the average diversification $d^{(2)}_i$ of countries
exporting the products exported by country $i$) are better able to capture the relevance of the country in the world trade than successive iterations. 
However, the FCM outperforms the MR (see Fig. \ref{fig:algorithm_comparison} for a visual comparison of algorithms' performance).
The FCM outperforms the elementary ranking of nodes by diversification [in 1996, $E_C(F)=0.357$ whereas $E_C(d)=0.332$]:
despite fitness and diversification are highly correlated, country fitness 
provides a better estimate of country importance with respect to diversification.
For products, the difference between the FCM and the MR is even larger (see Fig. \ref{fig:algorithm_comparison}).
Fig. \ref{fig:extinction} shows that ubiquity provides a poor estimate of node importance and it is outperformed by the MR scores $u^{(n)}$
with $n\geq 2$.
However, the original FCM ($\gamma=1$) performs remarkably better than the MR. This suggests that the extremality property of Eq. \eqref{metrics}
is an essential element to correctly rank products, which is confirmed by the monotonous dependence of $E_{P}$ on $\gamma$ for the FCM.
The generalized FCM with $\gamma>1$ outperforms the original algorithm ($\gamma=1$) in ranking nodes by importance
(see Figs. \ref{fig:extinction}, \ref{fig:algorithm_comparison}).

%

\subsection{Robustness of rankings against noise}
\label{robust}

Algorithm robustness against noise is a crucial property when studying incomplete or unreliable data \cite{ghoshal2011ranking,lu2011leaders}.
Data on international trade contains a substantial fraction of errors \cite{battiston2014metrics} which may arise from different sources.
In some cases information is missing: for instance, our dataset only contains data on the mutual exchanges between $72$ countries, while
exchanges between countries outside this core group are absent (see Materials and Methods and \cite{feenstra2005world}).
Moreover, original data sometimes are inaccurate and manually adjusted during the preparation of the dataset, 
and, in general, data can undergo various cleaning procedures.

This unreliability of data makes the robustness of an economic complexity metric with respect to noise a key property.
To study this robustness property, we use the same method of Ref. \cite{battiston2014metrics}: we randomly reverse a fraction $\eta$ of elements 
in the country-product matrix $M_{i\alpha}$ and, for each metric, we compute the Spearman's correlation between the rankings before and after the reversal.
Fig. \ref{fig:stability} shows this correlation for
three different values of $\eta$ ($\eta=0.01,0.05,0.1$), that are close to the expected level of noise in the world trade data which is $\eta\simeq 0.07$
(see Ref. \cite{battiston2014metrics} for further details).
The country ranking produced by the FCM has an optimal stability in the range $\gamma\in[1,1.5]$.
For products,  when $\gamma$ increases, 
$Q(\gamma)$ becomes more sensitive to the least-fit exporting country and, as a consequence,
the stability of the ranking according to $Q(\gamma)$ monotonously decreases. 
These findings imply that high values of $\gamma$ should not be used when the level of noise in data is high.

\subsection{Ranking volatility}

So far we have studied trade data from different years separately.
However, another interesting property of a ranking is its volatility, i.e. how the ranking changes with time.
In the absence of strong external factors such as a major war or a revolutionary scientific discovery, 
we expect fitness variations to be slow because
a country needs to develop new capabilities to grow its fitness, which is expected to be a time-consuming process lasting several years.
We study here the Spearman's correlation between the rankings produced by a method in two consecutive years.
Fig. \ref{fig:volatility} shows the results in the period 1996-1999.
For countries, the correlation $\rho(F(t),F(t+1))$ shows a maximum close to $\gamma=1$.
By contrast, for products $\rho(Q(t),Q(t+1))$
monotonously decreases with $\gamma$, which is another point against the use of large $\gamma$ values.
These results are in qualitative agreement with those obtained for stability against the random reversal of links (Fig. \ref{fig:stability}).
Moreover, these findings confirm that product complexity variations can be significant in the short-term \cite{cristelli2013measuring}.
This is due to the strong non-linear coupling between fitness $F$ and complexity $Q$: if a new country starts exporting a certain product,
the complexity of that product can change significantly.


\subsection{The effect of data incompleteness on the FCM}
\label{cleaning}

To understand the effect of data incompleteness on the
rankings produced by the FCM, we compute fitness and
complexity scores on a restricted dataset where
we include only the 72 core countries for which the complete information on mutual trade exchanges is available. While the idea of including only the
countries for which the full information is available may seem appealing, we find the results on the
restricted dataset misleading as, for example, some apparently simple products such as soy beans
appear at top positions of the product complexity ranking.
This happens because low­-fitness exporters of these products are not included in the restricted
dataset. In the case of soy beans, for example, only the United States, Brazil and Argentina exceed the RCA
threshold, which directly results in a high complexity score. In the 132-­countries dataset that we
used through the paper, there are $5$ other countries that exceed the RCA threshold, including Malawi
and Zambia, and the complexity score of soy beans is
correspondingly lower.
This example demonstrates that the information on the exports of low-­fitness countries is crucial for
the FCM because even a single low­-fitness country exporting
a certain product carries substantial information about
the product's complexity.
At the same time, the missing trade exchange information between the 60 non-­core countries is
comparatively less important because the non-core countries are typically low-fitness countries and the trades between them do not affect the fitness
of the most diversified countries.
Our choice to use the 132­-countries dataset is thus well justified.
The missing trades would be crucial only if we were interested in the import-export flows between two specific countries; however, 
this level of resolution is not needed for our work which focuses on methods that only require the binary matrix $M_{i\alpha}$ as input.

\section{Discussion}

To summarize, our article shows that the FCM is highly effective in ranking the nodes by importance 
in the country-product network. 
This property has already been observed in ecological networks \cite{dominguez2015ranking}
which indicates that the potential range of application of the FCM is broad and not restricted to economic systems.
Using the extinction area metric, we
provide a quantitative estimate of the performance gap between the FCM and the MR.
Our study of the generalized version of the FCM shows that, in general, the optimal choice of the extremality parameter $\gamma$ should be
based on a tradeoff
between the method's ability to capture the nestedness of the network and the ranking robustness against noise.
When the extremality parameter $\gamma$ is close to $1$ (i.e., close to the value set in the original metric defined by Eq. \eqref{metrics}),
the product ranking is determined both by product ubiquity and by the least-fit exporting country,
while the correlation between country fitness and diversification is maximal.
By choosing $\gamma$ around one, we maximize the correlation between the country ranking and
diversification, the robustness of the country ranking with respect
to noise and the correlation between the country rankings of two consecutive years.
When the extremality parameter $\gamma$ is large ($\gamma>2$), the product ranking becomes almost perfectly correlated with the ranking of products by
the fitness of the least-fit exporting country. As a consequence,
the rankings better capture the importance of nodes to the system stability but, at the same time, they are highly sensitive
to noise and volatile.
For this reason, we believe that the metric with large $\gamma$ should not be applied to noisy data, such as the world trade 
data, whereas it could still be informative
when applied to high-quality data.

Our findings deepen our understanding of the Fitness-Complexity method and have potential implications
for all the systems where the method has been applied, including
ecological networks \cite{dominguez2015ranking} and
the bipartite network of countries and academic fields \cite{cimini2014scientific}.
We stress that our findings are based on the evaluation of the rankings on the country-product networks
corresponding to different years.
The two methods has been also used to predict the future exports \cite{vidmer2015prediction} 
and the future economic growth of a country \cite{hausmann2014atlas,cristelli2015heterogeneous},
but which of the two metrics has more predictive power remains unclear.
The comparison between the studied network-based metrics with respect to their predictive power has not been addressed
in this work, and may constitute an interesting direction for future research on Economic Complexity.

\subsection*{Acknowledgements}

We thank Luciano Pietronero and Matthieu Cristelli for relevant suggestions for the preparation of the dataset and useful discussions
on the interpretation of our results.
This work was supported by the EU Project nr. 611272 GROWTHCOM.

\subsection*{Author contribution statement}

All authors designed the research; M.S.M. and A.V. wrote the code to perform the experiments;
all authors analyzed the data; M.S.M., M.M. and A.V. wrote the manuscript; all authors reviewed the manuscript.


%

%


%

\begin{thebibliography}{28}

\bibitem{ricardo1891principles}
D.~Ricardo, \emph{Principles of political economy and taxation} (G. Bell and
  sons, 1891)

\bibitem{romer1990endogenous}
P.M. Romer, Journal of Political Economy \textbf{98}(5 pt 2) (1990)

\bibitem{grossman1991quality}
G.M. Grossman, E.~Helpman, The Review of Economic Studies \textbf{58}(1), 43
  (1991)

\bibitem{hidalgo2009building}
C.A. Hidalgo, R.~Hausmann, Proceedings of the National Academy of Sciences
  \textbf{106}(26), 10570 (2009)

\bibitem{tacchella2012new}
A.~Tacchella, M.~Cristelli, G.~Caldarelli, A.~Gabrielli, L.~Pietronero,
  Scientific Reports \textbf{2} (2012)

\bibitem{brin1998anatomy}
S.~Brin, L.~Page, Computer Networks and ISDN Systems \textbf{30}(1), 107 (1998)

\bibitem{hausmann2014atlas}
R.~Hausmann, C.A. Hidalgo, \emph{The atlas of economic complexity: Mapping
  paths to prosperity} (MIT Press, 2014)

\bibitem{felipe2012product}
J.~Felipe, U.~Kumar, A.~Abdon, M.~Bacate, Structural Change and Economic
  Dynamics \textbf{23}(1), 36 (2012)

\bibitem{poncet2013export}
S.~Poncet, F.S. De~Waldemar, World Development \textbf{51}, 104 (2013)

\bibitem{cheng2013hidden}
Z.~Cheng, T.~Dongfeng, L.~Xiangqian, \emph{The hidden capabality network of
  product space}, in \emph{Service Operations and Logistics, and Informatics
  (SOLI), 2013 IEEE International Conference on} (IEEE, 2013), pp. 567--571

\bibitem{cristelli2013measuring}
M.~Cristelli, A.~Gabrielli, A.~Tacchella, G.~Caldarelli, L.~Pietronero, PLoS
  ONE \textbf{8}(8), e70726 (2013)

\bibitem{caldarelli2012network}
G.~Caldarelli, M.~Cristelli, A.~Gabrielli, L.~Pietronero, A.~Scala,
  A.~Tacchella, PLoS ONE \textbf{7}(10), e47278 (2012)

\bibitem{tacchella2013economic}
A.~Tacchella, M.~Cristelli, G.~Caldarelli, A.~Gabrielli, L.~Pietronero, Journal
  of Economic Dynamics and Control \textbf{37}(8), 1683 (2013)

\bibitem{battiston2014metrics}
F.~Battiston, M.~Cristelli, A.~Tacchella, L.~Pietronero, Complexity Economics
  \textbf{1}(1), 1 (2014)

\bibitem{allesina2009googling}
S.~Allesina, M.~Pascual, PLoS Computational Biology \textbf{5}(9), e1000494
  (2009)

\bibitem{dominguez2015ranking}
V.~Dom{\'\i}nguez-Garc{\'\i}a, M.A. Mu{\~n}oz, Scientific Reports \textbf{5}
  (2015)

\bibitem{feenstra2005world}
R.C. Feenstra, R.E. Lipsey, H.~Deng, A.C. Ma, H.~Mo, NBER Working Paper (11040)
  (2005)

\bibitem{hidalgo2007product}
C.A. Hidalgo, B.~Klinger, A.L. Barab{\'a}si, R.~Hausmann, Science
  \textbf{317}(5837), 482 (2007)

\bibitem{vidmer2015prediction}
A.~Vidmer, A.~Zeng, M.~Medo, Y.C. Zhang, Physica A: Statistical Mechanics and
  its Applications \textbf{436}, 188 (2015)

\bibitem{balassa1965trade}
B.~Balassa, The Manchester School \textbf{33}(2), 99 (1965)

\bibitem{cristelli2015heterogeneous}
M.~Cristelli, A.~Tacchella, L.~Pietronero, PLoS ONE \textbf{10}(2), e0117174
  (2015)

\bibitem{cimini2014scientific}
G.~Cimini, A.~Gabrielli, F.S. Labini, PLoS ONE \textbf{9}(12), e113470 (2014)

\bibitem{berkhin2005survey}
P.~Berkhin, Internet Mathematics \textbf{2}(1), 73 (2005)

\bibitem{pugliese2014convergence}
E.~Pugliese, A.~Zaccaria, L.~Pietronero, arXiv preprint arXiv:1410.0249  (2014)

\bibitem{bascompte2003nested}
J.~Bascompte, P.~Jordano, C.J. Meli{\'a}n, J.M. Olesen, Proceedings of the
  National Academy of Sciences \textbf{100}(16), 9383 (2003)

\bibitem{bascompte2010structure}
J.~Bascompte, Science \textbf{329}(5993), 765 (2010)

\bibitem{ghoshal2011ranking}
G.~Ghoshal, A.L. Barab{\'a}si, Nature Communications \textbf{2}, 394 (2011)

\bibitem{lu2011leaders}
L.~L{\"u}, Y.C. Zhang, C.H. Yeung, T.~Zhou, PLoS ONE \textbf{6}(6), e21202
  (2011)

\end{thebibliography}
%
\bibliographystyle{epj}

%
%

\end{document}